# Systematic literature review protocol

## *Identification and classification of feature modeling errors*

**Final version: 16/10/2020**

Dr. Samuel Sepúlveda, Dr. Jaime Díaz, Mg. Marcelo Esperguel
Dpto. Cs. de la Computación e Informática[1]
Centro de Estudios en Ingeniería de Software[2]
Universidad de La Frontera[3]
Temuco, Chile.

[1] http://dci.ufro.cl
[2] http://www.ceisufro.cl
[3] https://www.ufro.cl

## Abstract

***Context:*** The importance of feature modeling languages for software product lines and the planning stage for a systematic literature review.

***Objective:*** A protocol for carrying out a systematic literature review about the evidence for identifying and classifying the errors in feature modeling languages.

***Method:*** The definition of a protocol to conduct a systematic literature review according to the guidelines of B. Kitchenham.

***Results:*** A validated protocol to conduct a systematic literature review.

***Conclusions:*** A proposal for the protocol definition of a systematic literature review about the identification and classification of errors in feature modeling was built. Initial results show that the effects and results for solving these errors should be carried out.

***keywords:*** *feature modeling, systematic literature review, protocol, errors, identification, classification.*

## 1. Introduction

*Software Product Lines* (SPL) are considered an intensive set in software systems that share a common and managed set of *features* to satisfy the specific needs of a certain market segment, which is developed from a common set of fundamental elements and in a pre-established manner (Clements and Northrop 2003). In addition, the SPL as a discipline at both the academic and industrial levels is fully active and full of challenges, such as developing different types of software systems and new technologies, among others (Bashrosuh, Garba, et al. 2017).

SPL are builded in two stages: domain engineering and application engineering (Pohl, Bockle et al. 2005). In the domain engineering stage, the elements common to all SPL as well as those that vary from product to product are described. The application engineering stage is where the individual products of the SPL are built by reusing domain devices and exploiting the variability of the SPL. For the effects of this proposal, we focus on Domain Engineering, in particular in the sub-stage Domain Requirements Engineering.

One of the key concepts in SPL development is *variability*, which gives SPL the flexibility required to diversify and differentiate products (Galster, Weyns, et al. 2014). For example, a software product must be able to adapt to the needs of each client or allow options for configuration and specific *features*, so the products can reach different market segments (Pohl, Bockle, et al. 2005). As for domain engineering, it is common to describe SPL and manage their *variability* with the aid of a *feature model* (FM) (Asikainen, Mannisto, et al. 2006). FM was first presented as part of the FODA method (Kang, Cohen, et al. 1990). This model is still present, but with slight variations and adaptations for some SPL methods based on visual representations for the features of the product. The structure of a FM is a type of tree where its root node represents the product family and the *features* are organized throughout the tree. These *features* can be assembled to give rise to particular software products (Czarnecki and Wasowski 2007). One of the remarkable results from the work of (Heradio, Perez-Morago, et al. 2016) establishes that feature modeling has been the most important topic for SPL in recent years, having the best evolution behavior in terms of number of published papers and references used.

It must be considered that the use of FM involves a set of challenges, including in particular: (i) FM languages lack their own conceptual base, causing differences in the set of concepts included in each method that adopts FM, and (ii) current support and the level of maturity of the tools to build FM are quite limited. In this light, it may be noted that nowadays there are ambiguity issues with regard to FM, which result in redundancy problems, anomalies, inconsistency and mainly semantics issues (Sepúlveda, Cares, et al. 2012a; Sepúlveda, Cares, et al. 2012b).

All these considerations are in the foundations of the research project DI20-0060. The goal of this project considers to create a framework for solving the errors detected in the use of feature modeling languages. The framework, named FraSE-FML (acronym for Framework for Solving Errors in Feature Modeling Languages), considers a theoretical and an empirical perspective, including a set of formal models and algorithms to solve the errors detected for the modeling languages used in feature modeling for SPL. The expected result from this framework is the corrections of errors for the use of feature modeling languages. The impact with this proposal, beyond the generation of models without errors, is related to the development of better software products, this implies a lesser need for corrections and maintenance. The latter is relevant since these stages usually consume enough resources within the software development process.

The rest of this report is structured as follows. The next section describes the research method to follow, showing the details about the protocol definition for carrying out the systematic literature review. Finally, Section 3 presents the conclusions and future work.

## 2. Protocol definition

This section details the protocol definition of a systematic literature review (SLR). This protocol considers the three main phases according to the protocol defined by (Kitchenham 2004, Kitchenham and Charters 2007). These phases are: p*lanning, conducting, and reporting*.

### 2.1 Planning

The planning considers declaring the aim and need for the SLR, and defines the research questions (RQ) that the SLR must answer. Also, this considers the construction of the search string that will allow gathering the relevant papers for the study. Finally, the main considerations about the validation of this protocol are also considered.

#### 2.1.1 Aim and need

This SLR aims to summarize and synthesize the evidence about errors present in the modeling of features for SPL and the effects/impacts of these. See if there are already classifications associated with the types of errors and the solutions proposed.

The impact with this proposal, beyond the generation of models without errors, is related to the development of better software products, this implies a lesser need for corrections and maintenance. The latter is relevant since these stages usually consume enough resources within the software development process.

The motivation arises to propose a study that considers the common aspects and the deficiencies detected in the use of feature modeling languages and the tools implemented from these. The ambition of this proposal is to contribute to the SPL community, in particular for requirements engineering in the domain engineering stage, with a solution for well-designed models and with no presence of errors in the domain studied.

### 2.1.2 Research questions (RQs)

#### *2.1.2.1 Context:*

This SLR protocol is framed in the research project DI20-0060 that aims at the development of the framework FraSE-FML. The motivation that guides this proposal arises from the idea of avoiding the incorrect use of modeling languages for FM in SPL. In terms of the modeling of features in SPL, this proposal considers the common aspects and the deficiencies in syntax, semantics and semiotics detected in these modeling languages and the tools implemented from these. The ambition of this proposal is to contribute to the SPL community, in particular for the Domain Engineering with a solution for well-used feature modeling languages and with no presence of errors. It is possible to help the analysts (researchers and practitioners), to construct FM while avoiding the errors identified. At the industry level, the results of this proposal could be applied to build algorithms and develop tools to support the correct feature modeling.

The main idea consists of creating a framework (named FraSE-FML) that includes a set of formal models (meta modeling, semantics for modeling languages, and semiotic clarity) and algorithms that allow classify and solve the set of errors detected in the use of feature modeling languages.

#### *2.1.2.2 Research questions definition*

According to Kitchenham and Charters, we define a context for the RQs guiding this study (Kitchenham and Charters 2007). The context for the RQs arise from a more general question: *Will it be possible to identify and characterize the errors present in the use of feature modeling for SPL?*

The RQs that drive this SLR and their contribution to the general aim are shown in Table 1.

**Table 1.** Research questions & contribution.

| RQ# | Research Question | Aim | Possible answers or classification schema |
|---|---|---|---|
| RQ1 | What is the degree (level???) of validation for the feature modeling errors proposals?<br><br>RQ1.1 How many times the feature modeling errors proposals were used? | To determine how each proposal was validated.<br><br>To have an account for the times the proposal was used. | • quasi experiment<br>• case study<br>• data mining<br>• opinion survey<br>• lesson learnt<br>• example/proof of concept<br>• other<br>Adapted from (Kitchenham and Charters, 2013)<br><br>• 1, 2, 3, ... *n* times. |
| RQ2 | What kinds of errors do the feature modeling proposals solve? | To classify the type of errors solved as declared by each proposal. | Verbatim transcript from the paper. Open category. |
| RQ3 | What kind of benefits do the feature modeling proposals bring? | To determine the benefits declared by each proposal. | Verbatim transcript from the paper. Open category. |
| RQ4 | What are the technologies used for the feature modeling errors proposals? | To identify the technologies declared for each proposal. | • algorithm<br>• framework<br>• IDE/plugins/apps<br>• service/api<br>• taxonomy<br>• others (which ones?) |
| RQ5 | What is the origin of the feature modeling errors proposals? | To identify the origin of each proposal. | • Academia<br>• Industria<br>• Joint |
| RQ6 | What is the type of evidence collected for feature modeling errors proposals? (Research type?) | To determine the kind of each paper containing the proposal. | • Wieringa's classification (Wieringa et al., 2006) |

***i. Publication questions:*** additionally, a set of publication questions (PQs) has been included to complement the gathered information and characterize the bibliographic and demographic space. This includes the type of venue where the papers were published, amount of papers per year, and where the subject has been more developed. The details are shown in Table 2.

**Table 2.** Publication questions & contribution.

| PQ# | Publication Question | Aim | Possible answers or classification schema |
|---|---|---|---|
| PQ1 | What are the most relevant venues?<br>PQ1.1 Who published them? | To determine the distribution of papers by type of venue, i.e., journal, conference or workshop.<br>Besides, to determine what publishers are the most relevant for the studied subject. | • Journal (WoS & JCR quartiles, No WoS)<br>• Conference/Workshop, CORE ranking (or another)[4]<br>• Papers by publisher (graph bar) |
| PQ2 | How the quantity of papers has evolved across the time? | To determine the number of publications per year, in the period 2010-2020. | • Publication per year (graph bar).<br>• TimeLine |
| PQ3 | Which are the most active countries?<br>PQ3.1 What are the author's affiliations? | To determine the most active countries in the subject under study. Besides, to classify the author's affiliations in one of the two categories: academy or industry. This | • By Country<br>• By Center, University or Company |

[4] i.e. http://www.conferenceranks.com

| | | | |
|---|---|---|---|
| | | considers the affiliations of all authors as the same (don't matter if this is first, second, etc. or correspondence author). | |

#### 2.1.2.3 Search string definition

According the steps defined in (Kitchenham and Charters 2007):

- from the context & RQs ----> keywords
- keywords ---> synonyms
- build search string using PICOC (Petticrew and Roberts 2008)

See details in Table 3.

**Table 3.** PICOC criteria & details.

| Criteria | Scope | Detail in SE |
|---|---|---|
| ***Population*** | Who or What? | For SE should correspond to one of the following: (1) specific SE role, (2) a category of software engineer, (3) an application area or (4) an industry group. |
| ***Intervention*** | How? | Intervention in SE is defined as a methodology, tool, technology or procedure that addresses a specific issue. For example, performing specific tasks such as requirements specification, system testing, or software cost estimation. |
| ***Comparison*** | Compared to what / what is the alternative? | The comparison element is not applicable to our RQs, since they did not involve the comparison of the collected papers against any commonly used language or technique (the control condition). |
| ***Outcomes*** | What are we trying to accomplish, improve, effect? | Outcomes should relate to factors of importance to practitioners such as improved reliability, reduced production costs, and reduced time to market. |
| ***Context*** | Under what circumstances? | For Software Engineering, this is the context in which the comparison takes place (e.g. academia or industry), the participants taking part in the study (e.g. practitioners, academics, consultants, students), and the tasks being performed (e.g. small scale, large scale). |

- ***Keywords***
  - feature
  - model
  - error
  - software
  - product
  - line
  - identify
  - characterize
- ***Synonyms***
  - Main Context
    - feature model / modelling
    - feature diagram
    - software product family

- software product lines
    - Error
        - error
        - mistake
        - inconsistency
        - anomaly
        - failure
    - Process
        - identification
        - formalization
        - classification
        - checking
        - validation
        - analysis
        - model checking/ validation / verification / querying

- ***PICOC***
    - *population*: in our case, an application area was considered, in particular feature modeling for Software Product Lines.
    - *intervention*: in our case, the intervention is part of a methodology, in particular Requirements Engineering, Domain Engineering and Modeling Language.
    - *comparison*: N/A (there is no “control condition).
    - *outcomes*: in our case the main outcomes of the RQs are the origin, validation, formalization and level of adoption in the academy or industry for the FM solving errors proposals.
    - *context*: in our case the proposals can take place in academia or industry, being developed by academics or practitioners, solving or treating FM errors.

**i. Search string**

( "feature modeling" OR “feature diagram” OR “software product line” ) AND
( "error" OR “mistake” OR “inconsistency” OR “anomaly” OR “failure” OR “problem”) AND (“identification” OR “formalization” OR “classification” OR “checking” OR “validation” OR “analysis” OR “verification” OR “solution”)

### ii. Background

As a test for calibration of the search string, we carried out an initial search as follows:

- Date test: 01/10/20
- Source: Scholar Google[5]
- Time-span: 2010-2020
- Excluding: cites and patents
- Outcome: 5010 records

**iii. Checking relevance:**

- Considering the first 10 pages, TOTAL 100 records (2% total sample).
- From the titles, it is observed that for the first 10 pages there are: 0-2-4-2-4-4-5-4-4-3 number of non-relevant works on the subject respectively.
- Then, to verify the precision of the search string, we use the Precision concept (P) as a useful measure of relevancy. This is defined as the number of true positives (Tp), over the number of true positives plus the number of false positives (Fp).

[5] http://www.scholargoogle.com

- P = Tp / (Tp+Fp).

The details for the 10 pages returned by Scholar Google are shown in Table 4.

**Table 4.** Precision for the results returned in Scholar Google using the search string.

| Page range | Tp | Fp | P |
|---|---|---|---|
| 1st page | 10 | 0 | 100% |
| 1st-2nd page | 18 | 2 | 90% |
| 1st-3rd page | 24 | 6 | 80% |
| 1st-4th page | 32 | 8 | 80% |
| 1st-5th page | 38 | 12 | 76% |
| 1st-6th page | 44 | 16 | 73% |
| 1st-7th page | 49 | 21 | 70% |
| 1st-8th page | 55 | 25 | 69% |
| 1st-9th page | 61 | 29 | 68% |
| 1st-10th page | 68 | 32 | 68% |

Then, the average precision for the first 10 pages is $P_{average}$ = 78%. We considered this value as good, indicating that approx. 8/10 papers selected automatically are relevant for the study.

#### 2.1.2.4 Protocol validation

Initially, one researcher builds the protocol, then, the rest of researchers will must evaluate and discuss the correctness and completeness for the protocol.

Next, the changes and corrections will be agreed among the researchers and a validated version of the protocol will be compiled.

According to (Kitchenham and Charters 2007), we pretend to evaluate the consistency of the protocol. To do this, we have to answer these questions:

- Are the search strings appropriately derived from the RQs?
- Will the data to be extracted properly address the RQs?
- Is the data analysis procedure appropriate to answer the RQs?

Ideally, the agreed compiled version of the protocol will be sended to an external expert to review it.

### 2.2. Conducting

The conducting phase considers to define the search strategy, the inclusion/exclusion criteria, and the data extraction process. Also, it considers a quality assessment about the gathered evidence and an initial analysis of the threats to validity of the results.

#### 2.2.1 Search strategy

Since the defined data sources include search engines, the strings will be entered sequentially with the combinations of these and adapting it to each search engine as appropriate.

Data sources: according to (Brereton and Kitchenham 2007, Kitchenham and Charters 2007) we consider the sources detailed in Table 5, that are recognized among the most relevant in the SE community.

**Table 5.** Data sources.

| Fuente | Link |
|---|---|
| ACM Digital Library | https://dl.acm.org |
| IEEE Xplore | https://ieeexplore.ieee.org/Xplore/home.jsp |
| IET Digital Library | https://digital-library.theiet.org/content/journals |
| Science Direct | https://www.sciencedirect.com |
| Springer Link | https://www.springer.com/ |
| Wiley Inter Science | https://www.onlinelibrary.wiley.com/search/advanced |

***Time period:*** the search considers the period 2010-2020. This is because in preliminary searches using Scholar Google, we can see that approx. more than 80% of results concentrate in this period. Besides, a relevant SLR carried out for Benavides in 2009 includes the main concepts for variability modeling in 20 years (Benavides et al., 2009).

***Language***: the selected language will be English. It is considered that this decision, in general, would not become a bias for the SLR, because even though there may be references in other languages that could be discarded, many of these works, given their relevance, will also be published in English, in high impact magazines or conferences.

***Search process:*** this process considers two steps.

- Step 1: automatic search on selected electronic databases.
- Step 2: snowballing process (backward & forward), starting from the final list of selected papers to complement the automatic search (Wohlin 2014).

From the collected papers in step 1, we must consider:

- Different versions of the same proposal (expanded works or versions), the criterion will be: keep the last one.
- Search and eliminate for duplicated works.

For the remaining papers, the researchers will independently review the assigned papers and will decide if these are relevant or not for the SLR only reviewing title, abstract and keywords for the papers. The set of remaining papers will be filtered applying the exclusion criteria (EC). The definition of every EC is shown in Table 6.

**Table 6.** Exclusion criteria & details.

| EC# | Description |
|---|---|
| EC1 | The paper is not written in English. |
| EC2 | The paper is not peer reviewed (Posters, conference proceedings, books/book chapter, tutorials, slides, PhD or master thesis and any piece of work that can be considered as grey literature) |
| EC3 | The paper is a secondary study (they, if any, will be considered in our Related works section). |
| EC4 | The paper is a short paper (four or less pages). |
| EC5 | The focus of the paper is not on proposals treating errors in FM. |
| EC6 | The paper does not show empirical results, at least at the toy example level. |

### 2.2.2 Resolving differences and avoiding bias

- Collaborative definition of the SLR protocol and RQs.
- External validation for the search string
- Publication of the protocol for public scrutiny on the arXiv platform
- To avoid any potential bias due to a particular researcher examining each paper, we verified that the manner of applying and understanding the exclusion criteria was similar for the researchers involved in the SLR (inter-rater agreement).
- The researchers individually decide on the inclusion/exclusion of a set of xx papers randomly chosen from those retrieved by a pilot selection.
- A test of concordance based on the *Fleiss' Kappa statistic* will be performed as a means of validation (Gwet 2002).
    - if *Kappa* ≥ 0.75 then *the criteria is clear enough (Fleiss 1981).*
    - else, *the criterion must be reviewed to get accordance in its interpretation and application*.

Another way to do this task is following the criteria defined by (Petersen, Vakkalanka et al. 2015).

With the final list of selected papers, one round of snowballing process (backward & forward) will be carried out using these papers as "seeds" (Wohlin 2014).

### 2.2.3 Data extraction

For the selected papers, relevant data will be extracted in order to answer the RQs and PQs.

The meta-data collected for each paper: (i) title, (ii) authors (each of them), (iii) publication year, (iv) type of publication and ranking, (v) tools reported, and (vi) results and future work.

Besides, the detailed data for answering each RQ must be registered. All this data will be consolidated in a spreadsheet as detailed in Table 7. The details for the PQ are shown in Table 8.

**Table 7.** RQ - data extraction form.

| **Paper#** | **RQ1** | **Evidence RQ1** | **RQ2** | **Ev RQ2** | **….** | **RQ5** | **Ev RQ5** | **RQ6** | **Ev RQ6** |
|---|---|---|---|---|---|---|---|---|---|
| 01 | | | | | …. | | | | |
| 02 | | | | | …. | | | | |
| ... | ... | ... | ... | | …. | | | | ... |
| nn | | | | | …. | | | | |
| **Values** | type of venue (text) | | type of experien ce (text) | | …. | type of actor and output (text) | | ... | |
| **Chart** | Bar | | Bar | | …. | Bar | | ... | |

**Table 8.** PQ - data extraction form.

| **Paper#** | **PQ1** | **PQ2** | **PQ3** |
|---|---|---|---|
| 01 | | | |
| 02 | | | |
| ... | | ... | .... |
| nn | | | |
| **Values** | type of study and publisher (text) | year of publication (integer) | country names (text) |
| **Chart** | Bar | Line | Bar |

### 2.2.4 Graphical summary

- Another resource to consolidate data is using a weighted word cloud from the abstracts of the selected paper to show the most relevant concepts. See Fig. 1 as an example.

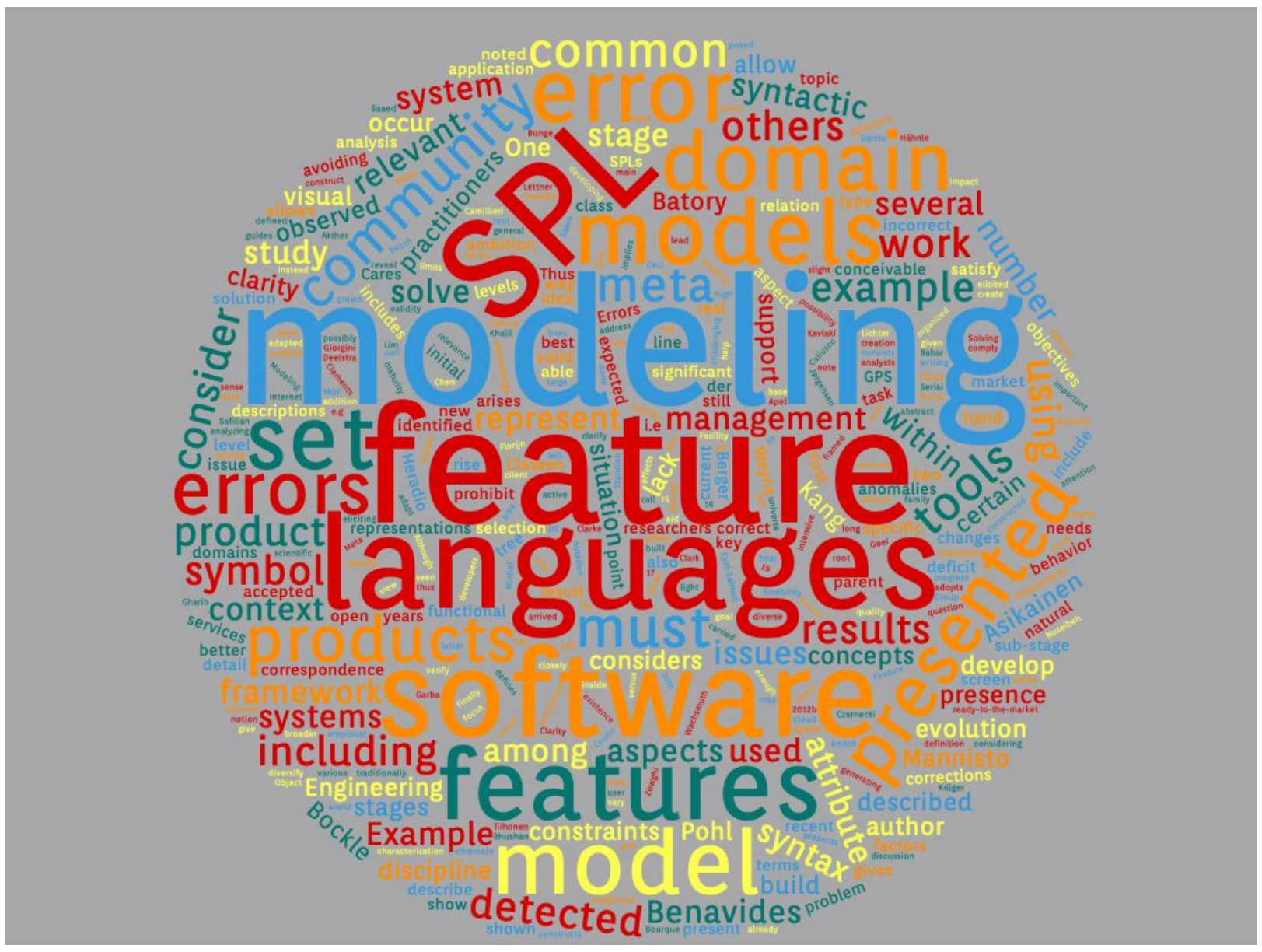

**Fig. 1** Cloud tag SPL keywords as example.

- To see the relevant concepts, a tool for seek relations in the text will be used. See Figs. 2-3 as an example of using Termine[6].

[6] http://www.nactem.ac.uk/software/termine/

Found **18** terms in 8.16 seconds - all terms (in table) (in text) - threshold: 0 Apply

H1 : It is possible to identify and formalize the errors present in the use of feature modeling for SPL , considering elements from the analysis of the language. To identify a set of errors in the use of FM languages for SPL , this proposal will endeavor to conduct an exhaustive review of the state of the art around feature modeling in SPL , for which a systematic literature review will be performed ( Kitchenham 2004 ). . The motivation arises to propose a study regarding the feature modeling in SPL that considers the common aspects and the deficiencies in syntax , semantics and semiotic clarity detected in the use of these modeling languages and the tools implemented from these The ambition of this proposal is to contribute to the SPL community , in particular for requirements engineering in domain engineering stage , with a solution for well- designed models and with no presence of syntactic , semantic or semiotic errors in the domain being studied. The goal of this research proposal is to create a framework for solving the errors detected in the use of feature modeling languages The hypotheses and objectives are presented in Section 2. The framework , named FraSE-FML ( acronym for Framework for Solving Errors in Feature Modeling Languages ) , considers a theoretical and an empirical perspective , including a set of formal models and algorithms to solve the errors detected for the modeling languages used in feature modeling for SPL The description of this framework is presented in Section 3. The expected result from this proposal is the corrections of errors for the use of feature modeling languages. The impact with this proposal , beyond the generation of models without errors , is related to the development of better software products , this implies a lesser need for corrections and maintenance The latter is relevant since these stages usually consume enough resources within the software development process. . The motivation that guides this proposal arises from the idea of avoiding the incorrect use of modeling languages for FM in SPL In terms of the modeling of features in SPL , this proposal considers the common aspects and the deficiencies in syntax , semantics and semiotics detected in these modeling languages and the tools implemented from these The ambition of this proposal is to contribute to the SPL community , in particular for the stage of RE in Domain Engineering with a solution for well-used feature modeling languages and with no presence of errors It is possible to help the analysts ( researchers and practitioners ) , to construct FM while avoiding the errors identified At the industry level , the results of this proposal could be applied to build algorithms and develop tools to support the correct feature modeling. .

**Fig. 2** Cloud tag for keywords from the abstracts.

| Rank | Term | Score |
|---|---|---|
| 1 | feature modeling | 6.6 |
| 2 | modeling language | 5.5 |
| 3 | feature modeling language | 3.169925 |
| 4 | spl community | 2 |
| 4 | domain engineering | 2 |
| 4 | common aspect | 2 |
| 4 | well-used feature modeling language | 2 |
| 8 | systematic literature review | 1.584962 |
| 8 | feature modeling languages | 1.584962 |
| 10 | semiotic clarity | 1 |
| 10 | empirical perspective | 1 |
| 10 | software product | 1 |
| 10 | exhaustive review | 1 |
| 10 | semiotic error | 1 |
| 10 | formal model | 1 |
| 10 | solving errors | 1 |
| 10 | fm language | 1 |
| 18 | software development process | 0 |

**Fig.3** Cloud tag for keywords from the abstracts.

### 2.2.4 Quality assessment (QA)

In order to address bias, external and external validity, each selected paper was subjected to a QA (Kitchenham and Charters 2007). To answer about the QA we will use the criteria proposed by (Dyba and Dingsoyr 2008). These criteria are summarized in five questions that can be answered with *yes/partially/no*. Other papers have already adopted the same criteria, among these are (Chen and Babar 2011, Unterkalmsteiner, Gorschek et al. 2012, Sepúlveda, Cravero et al. 2015). The QA is summarized in Table 9.

**Table 9.** QA questions.

| QA# | QA question | Yes | Partially | No |
|---|---|---|---|---|
| QA1 | Is the aim of the research sufficiently explained? | ## (%) | ## (%) | ## (%) |
| QA2 | Is the paper based on research methodology? | ## (%) | ## (%) | ## (%) |
| QA3 | Is there an adequate description of the context in which the research was carried out? | ## (%) | ## (%) | ## (%) |
| QA4 | Are threats to validity taken into consideration? | ## (%) | ## (%) | ## (%) |
| QA5 | Is there a clear statement of findings? | ## (%) | ## (%) | ## (%) |

### 2.2.5 Threats to validity

This sub-phase considers the types of validity defined by (Petersen et al., 2013; Ampatzoglou et al, 2018).

- Theoretical validity.
- Descriptive validity.
- Interpretive validity.
- Generalizability.
- Reliability.

In a complementary way, as a quality indicator for the report of this SLR, its structure will be validated using the PRISMA Statement[7] (Preferred Reporting Items for Systematic Reviews and Meta-Analyses).

## 2.3. Reporting

To state the relevance of this SLR, the reporting phase considers two steps.

- Step1: publish the reviewed SLR protocol at Arxiv platform[8].
- Step2: publish the SLR results in JCR journal. Initially, we consider the following options: IST[9], JSS[10], SoSym[11], Software evolution and process, etc.

Adapting the structure recommended by (Kitchenham 2004). The main sections considered are:

1. Introduction
   a. context
   b. motivation
   c. aim
   d. need and
   e. structure of the paper.
2. Background, main concepts about:
   a. SPL
   b. Variability modeling
   c. FM
   d. Errors using FM
   e. Syntactic, semantic and semiotic concepts.
3. Methodology: explain step by step how the protocol will be carried out.
4. Results and discussion
   a. main findings and results

[7] http://www.prisma- statement.org
[8] https://arxiv.org
[9] https://www.journals.elsevier.com/information-and-software-technology
[10] https://www.journals.elsevier.com/journal-of-systems-and-software
[11] https://www.springer.com/journal/10270

b. answers to RQs and PQs
    c. QA analysis and
    d. threats validity to the paper
5. Related work: summary of the main related work and RQs answered.
6. Conclusions
    a. conclusions and
    b. further lines of research.
7. Acknowledgements: recognize the support for the paper.
8. References:
    a. list of bibliographic references
    b. list of selected papers.
9. Appendix: detailed information (tables and graphs) for complementing data for sections 3, 4, 5, and 6.

# 3. Conclusions and future work

We presented a proposal for the protocol definition of a SLR to summarize and synthesize the evidence about the errors present in the modeling of features for SPL and the effects/impacts of these.

The initials results show that a more detailed review of the effects and results for solving these errors should be carried out.

As future work we pretend to be deep in the analysis of the errors in featuring modeling, considering the definition of a framework to face this problem.

## *Acknowledgments*

Samuel Sepúlveda thanks to Research Project DI20-0060 supported by Vicerrectoría de Investigación y Postgrado, Universidad de La Frontera. Thanks to Oscar Aguayo for his useful technical support.